\documentclass[12pt]{article}
\usepackage{amsmath}
\usepackage{amssymb}

\usepackage{graphicx}
\usepackage{latexsym}
\usepackage{bm}
\usepackage{color}
\usepackage{enumitem}
\usepackage[toc,page]{appendix}
\usepackage{mathtools}
\usepackage{amsmath}
\usepackage{slashed}
\usepackage{hyperref}
\usepackage{caption}
\usepackage{subcaption}
\usepackage{float}

\usepackage{tikz}
\usetikzlibrary{decorations.pathmorphing}
\numberwithin{equation}{section}

\tikzset{snake it/.style={decorate, decoration=snake}}

\begin{document}

\begin{titlepage}
    \setcounter{page}{1} \baselineskip=15.5pt \thispagestyle{empty}
    
    \begin{flushright}
		{\footnotesize{SLAC-PUB-17149, SU-ITP-17/09}} \date{}
	\end{flushright}
	
	\vspace{0.1cm}
	\begin{center}
		{\fontsize{21}{34}\selectfont \sc
		Hyperbolic Black Holes and  \\[.3cm]
		Open String Production
		\vspace{4mm}
		}
	\end{center}
	
	\begin{center}
		{\fontsize{13}{30}\selectfont Danjie Wenren}
	\end{center}
	
	\begin{center}

		\textsl{Stanford Institute for Theoretical Physics, Stanford University, Stanford, CA 94305, USA and}

		\textsl{ SLAC National Accelerator Laboratory, 2575 Sand Hill Rd., Menlo Park, CA 94025, USA}
\end{center}
\vspace{.5cm}
{
\begin{center}
\textbf{Abstract} 
\end{center}
\noindent In this paper we investigate open string pair production effects near hyperbolic black holes in AdS$_5$. We study the classical dynamics of D3 probe branes in this background and their quantum tunneling rate across the horizon from the inside using a method similar to that of Kraus, Parikh and Wilczek. The hyperbolic black holes can decay through these tunneling events and their lifetime is estimated. The radiated branes move towards the boundary and do not bounce back. Their world-lines do not intersect directly and they do not hit the singularity either, providing a clean scenario for studying the non-adiabatic pair production effects of open strings stretched between them near horizon. We find that there is a well defined parameter regime where there can be significant pair production. This requires the radiated branes to be highly boosted. However, the radiation of such branes still has small back-reaction on the black hole background, though the open string pair production on them can potentially alter the background. We comment on the possible relation of our model and the AMPS paradox. Some issues and future directions are discussed in the end.
\vspace{0.3cm}
 }
	
\end{titlepage}

\section{Introduction} \label{section:intro}
String theory predictions for processes involving large boost are oftentimes beyond the naive extrapolation of effective field theory (EFT). One of such examples is demonstrated in \cite{Silverstein:2014yza}, where it is shown that there can be a significant amount of open string production between two successive D-branes dropped in from infinity to a Schwarzschild black hole. The large time separation between the two branes causes large relative boost between the two branes at horizon crossing, which further leads to large non-adiabaticity of the frequencies of the open strings stretched between them. Using a first quantized method developed therein, \cite{Silverstein:2014yza} shows that in a well defined window of kinematic parameters, there can be unsuppressed open string production and the effect can be further enhanced by the large relative boost. These produced open string could later decay through other string-theoretic processes, resulting in a state well above the EFT vacuum.

Similar ideas are also explored in some other works, such as \cite{Puhm:2016sxj, Dodelson:2015toa, Dodelson:2017hyu, Dodelson:2017emn}. In particular \cite{Dodelson:2017hyu} shows that the large relative boost can lead to interesting phenomena in string scattering amplitude. It calculates gauge invariant six-point S-matrix element, convolved with the appropriate wave packets so that the scattering on wavepacket peak can only happen due to direct interaction via  string spreading effects. It is then shown that the scattering amplitude has support consistent with longitudinal spreading as large as $\alpha' E$ with $E$ being the energy scale of the scattering. This is well beyond those predicted by the analogous EFT and passes strong tests that scattering happens at peak, not on tail of wavepackets. Other stringy effects are also discussed extensively in literature, such as Bremsstrahlung \cite{McAllister:2004gd, Bachlechner:2013fja} and inelastic effects \cite{Amati:1987uf, Amati:1987wq, Veneziano:2004er, DAppollonio:2013okd, Giddings:2006vu, Giddings:2007bw}.

These studies are particularly interesting in the light of the recent AMPS paradox \cite{Almheiri:2012rt, Almheiri:2013hfa}, which suggests inconsistency between local effective field theory and the full underlying quantum theory of gravity in the presence of black holes. Various attempts have been being made in literature to resolve the paradox, such as those in the context of the AdS/CFT correspondence and quantum information theory. However, potential resolutions from dynamical effects in string theory have been relatively less discussed. Given the various distinctions between the predictions of string theory and effective field theory mentioned above, it seems natural and important to further study the non-adiabatic effects in string theory and understand the breakdown of EFT in this context.

In this work we consider open string production effects near hyperbolic black holes \cite{Emparan:1999gf} in Anti-de Sitter space (AdS). In particular we set our analysis in asymptotic AdS$_5\times S^5$, which has been extensively studied in the context of the AdS/CFT correspondence. These black holes are the higher dimension generalizations of BTZ black holes \cite{Banados:1992wn, Banados:1992gq} and can be thought of as formed by collapsing a stack of D3 branes. As we will see in Section \ref{section:setup} there is a continuous variable that determines the size of the black hole relative to the radius of AdS. A nice feature of this family of black holes is that one member of it is isometric to pure AdS, making it a relatively easier scenario for studying its horizon and interior \cite{Horowitz:2009wm}. This can be understood in a similar way as the relation between BTZ black hole and pure AdS$_3$, that is, this particular hyperbolic black hole can be constructed from the Poincar\'e patch of pure AdS$_5$ by compactifying the transverse 3-space $\mathbb{H}^3$ using an appropriate discrete subgroup. We will use this fact as a check for our open string production and no production results.

As with other large black holes in AdS, the hyperbolic black holes we consider are in thermal equilibrium with their Hawking radiation, making them especially long-lived as compared with black holes with open asymptotic boundaries. However, they can still decay by radiating D3 branes through quantum tunneling effects. We will compute the probability of such tunneling events and estimate the time separation between two consecutive brane emission events. The lifetime of such black holes is then estimated. The time line of brane emission and black hole lifetime seem to agree with that of the advent of the putative firewall in the AMPS paradox.

The mechanism of brane radiation provides a scenario where there are naturally branes lined up near horizon and separated by large distance in time. It is interesting to study the various non-adiabatic effects for these branes. In particular we consider open string production effects between two consecutively produced branes, which are roughly separated by a time that scales as $e^{CN^p}$, where $N$ is the number of branes inside the black hole and $p$ and $C$ constants. This large time separation can act as an enhancement factor for the open string production effects as one would expect from previous works \cite{Silverstein:2014yza, Puhm:2016sxj}. To get significant production, the radiated branes need to take away a fair amount of energy\footnote{Note that this is the total energy as defined by the Hamiltonian, not the kinetic energy which can blueshifted near the horizon. It will be precised defined in Section \ref{section:brane:radiation}.} so that there can be a large relative boost between them, even though that energy is still parametrically small compared to that of the black hole, leading to small back-reaction.

We will estimate the open string production level near horizon using the real time method developed in \cite{Silverstein:2014yza} and further discussed in \cite{Puhm:2016sxj}. For each oscillation level $n$ of the open string spectrum, the produced string density is estimated using the non-adiabaticity figure of merit $\dot{\omega}/\omega^2$ as $e^{-\omega^2/\dot{\omega}}$, where $\omega$ is the frequency of the open string and dot represents the time derivative. The total amount of production is computed by summing over all the oscillation levels and folding in the Hagedorn density of string state $e^{\sqrt{8\pi^2 n}}$.

The rest of the paper is organized as follows. Section \ref{section:setup} introduces in more detail about the black hole background under consideration and the coordinate systems we will use. Section \ref{section:brane:radiation} studies the classical dynamics of probe branes in this background and their quantum tunneling effects across the horizon. The brane radiation time scale and black hole lifetime are then estimated. Section \ref{section:open:string} discusses the non-adiabatic production of open strings stretched between the radiated branes. A brief summary and discussion about future directions are given in Section \ref{section:summary}.

\section{Setup} \label{section:setup}
The black hole solutions we consider in this paper are the hyperbolic black holes in AdS$_5$
\begin{equation}
	ds^2 = -f(r) dt_s^2 + \frac{dr^2}{f(r)} + r^2 d\sigma^2
	\label{equation:schwarzschild:metric}
\end{equation}
with 
\begin{equation}
	f(r) = \frac{r^2}{l^2} - 1 - \frac{\mu}{r^2},
\end{equation}
where $l$ is the AdS radius and $\mu$ is a continuous parameter that determines the relation between the black hole radius and $l$. This type of black hole has hyperbolic horizon with $d\sigma^2$ denoting the metric for a unit hyperbola $\mathbb{H}^3$. For the solution to be a black hole, the $\mathbb{H}^3$ directions are identified under its appropriate discrete subgroup, though the exact form of $d\sigma^2$ or the subgroup used is not important for our purpose. Let us denote the compactified $\mathbb{H}^3$ by $\Sigma$. These solutions can also be understood from the perspective of the AdS/CFT correspondence where the full background is AdS$_5\times S^5$ and $l$ is fixed via flux compactification as
\begin{equation}
	\label{equation:l:in:N}
	l^4 = 4\pi g_s\alpha'^2 N,
\end{equation}
where $N$ is the number of D3 branes used to form the AdS$_5\times S^5$ geometry and $g_s$ is the string coupling. The D3 branes wrap around the $\mathbb{H}^3$ directions and are evenly distributed along the $S^5$ directions. In this work we will consider the process of such black holes radiating D3 branes through quantum tunneling, and open string production on pairs of them.

The black hole radius is determined by the larger root of $f(r)$ and the result is 
\begin{equation}
	r_h = \frac{l}{\sqrt{2}}\left(1 + \sqrt{1+4\frac{\mu}{l^2}}\right)^{1/2}.
\end{equation}
This requires that in order for (\ref{equation:schwarzschild:metric}) to be a valid black hole, we need the following two equivalent conditions
\begin{equation}
	\mu \geq -\frac{l^2}{4}, \quad r_h \geq \frac{l}{\sqrt{2}}.
\end{equation}
The equivalence of the two conditions is in the sense that $r_h$ can act as a parameter for this type of black holes as much as $\mu$ can. Later when we compute the brane radiation rate and the open string production rate we will use $r_h$ since it is more convenient that $\mu$. The thermodynamics of these black holes is studied in \cite{Emparan:1999gf}, where the temperature and mass are found to be
\begin{align}
	\label{equation	:temperature}
	T &= \frac{f'(r_h)}{4\pi} = \frac{2r_h^2 - l^2}{2\pi l^2 r_h} \\%
	\label{equation:black:hole:mass}
	M &= \frac{3V_s}{16\pi G_5}\left(\mu + \frac{l^2}{4}\right) = \frac{3V_s}{8\pi^2\left(4\pi g_s\right)^{1/4}}\frac{N^{\frac{7}{4}}}{\sqrt{\alpha'}}\left(\frac{r_h^2}{l^2} - \frac{1}{2}\right)^2,
\end{align}
with $V_s$ being the volume of the unit $\Sigma$ and $G_5$ being the five-dimensional Newton's constant. To get the last equation of (\ref{equation:black:hole:mass}) we have used the following equation from the standard AdS/CFT dictionary
\begin{equation}
	N^2 = \frac{\pi l^3}{2G_5}.
\end{equation}

Perhaps a more useful coordinate system is the Gullstrand-Painlev\'e coordinates
\begin{align}
	ds^2 &= -\frac{r^2}{l^2}dt_g^2 + \frac{l^2}{r^2}\left(dr + \frac{r}{l}\sqrt{1+\frac{\mu}{r^2}}dt_g\right)^2  + r^2d\sigma^2 \nonumber \\%
		&= -\left(\frac{r^2}{l^2} - 1- \frac{\mu}{r^2}\right)dt_g^2 + 2\frac{l}{r}\sqrt{1+ \frac{\mu}{r^2}}dt_g dr + \frac{l^2}{r^2}dr^2 + r^2 d\sigma^2,
		\label{equation:painleve}
\end{align}
which can be obtained from (\ref{equation:schwarzschild:metric}) by a temporal reparametrization
\begin{equation}
	dt_s = dt_g - \frac{l}{rf(r)}\sqrt{1+\frac{\mu}{r^2}}dr.
\end{equation}
Compared to the Schwarzschild metric (\ref{equation:schwarzschild:metric}), the Gullstrand-Painlev\'e metric maintains explicit time translational symmetry and does not have poles at horizon and thus making it more suitable for studying processes crossing the horizon. In fact this time coordinate $t_g$ can be shown to be conformal to the proper time of an auxiliary D3 brane observer falling into the black hole from the boundary\footnote{This is the kind of in-falling observer with conserved energy $\omega = 0$, as made more precise in Section \ref{section:brane:radiation}. (\ref{equation:momentum:infalling:in:r}) then shows $p = 0$, which leads to the equation of motion
\begin{equation}
	\frac{dr}{dt_g} + \frac{r}{l}\sqrt{1+\frac{\mu}{r^2}} = 0
\end{equation}
as seen from (\ref{equation:conjugate:momentum}). The proper time of this observer can be computed as
\begin{equation}
	d\tau^2 = \frac{r^2}{l^2}dt_g^2\left(1 - \frac{l^4}{r^4}\left(\frac{dr}{dt_g}+\frac{r}{l}\sqrt{1+\frac{\mu}{r^2}}\right)^2\right) = \frac{r^2}{l^2}dt_g^2,
\end{equation}
which is conformal to $dt_g^2$.
}. 
One might be concerned that the large relative boost between such auxiliary observer and the produced branes may lead to open string production artifacts, which are not directly related to the non-adiabaticity from the relative motion of the radiated branes. This, however, is not a problem for our calculation because we will estimate the non-adiabaticity using the proper time of the radiated branes. Different from previous works, our analysis needs no brane to be dropped in from infinity.

 Another nice coordinate system is proposed in \cite{Puhm:2016sxj} where the effect from singularity can be excluded. Indeed for the scenarios where the string-theoretic production is catalyzed by the branes dropped in from the boundary, it is crucial to eliminate such effect, from which large non-adiabatic production is expected. In our work, however, the radiated branes start in the close vicinity of the horizon and move out towards the boundary. Because of their Ramond-Ramond charge, they also do not turn back to the horizon. Hence their world-volumes cannot be extended backward to the singularity due to the fact that classically nothing can cross the horizon from the inside. For these reasons we will work with the Gullstrand-Painlev\'e coordinates in this paper.
 
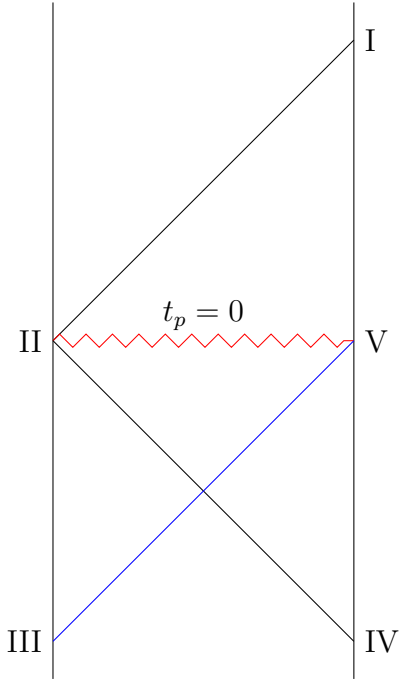
\begin{figure}
	\centering
	\begin{tikzpicture}
		\draw[-] (2, -4.5) -- (2, 4.5);
		\draw[-]  (-2, -4.5) -- (-2, 4.5);
		
		\node (I)		at (2,4)[right] {I};
		\node (II)		at (-2, 0)[left] {II};
		\node (III) 	at (-2, -4)[left] {III};
		\node (IV) 	at (2, -4)[right] {IV};
		\node (V) 	at (2, 0)[right] {V};
		
		\draw[-] (2,4) -- (-2, 0);
		\draw[decorate, decoration=zigzag, red] (-2,0) --  (2,0);
		\draw[-] (-2, 0) -- (2, -4);
		\draw[-, blue] (2,0) -- (-2, -4);
		\draw node[above] (0,0) {$t_p = 0$};
	\end{tikzpicture}
	\caption{Penrose diagram of the Poincar\'e patch and the corresponding hyperbolic black hole with $\mu = 0$. The former covers the region formed by point I, II and IV while the latter only covers the lower half of it. Similar to BTZ black holes, the singularity at $t_p = 0$ is formed due to the compactification of the $\mathbb{H}^3$ directions. The event horizon is shown in blue.}
	\label{fig:penrose:poincare:patch}
\end{figure}

One special member of this type of black holes is the one with $\mu = 0$, which is isometric to pure AdS$_5$. The explicit coordinate transformation between its Gullstrand-Painlev\'e coordinates and Poincar\'e coordinates is
\begin{equation}
	\label{equation:GP:2:Poincare}
	r = -\frac{r_p t_p}{l}, \quad t_g = -l \log\left(-\frac{t_p}{l}\right)
\end{equation}
or conversely
\begin{equation}
	t_p = -l \exp\left(-\frac{t_g}{l}\right), \quad r_p = r \exp\left(\frac{t_g}{l}\right).
\end{equation}
This transformation leads to the metric of the corresponding Poincar\'e patch
\begin{equation}
	\label{equation:poincare:patch}
	ds^2 = \frac{r^2_p}{l^2}\left(-dt_p^2 + t_p^2 d\sigma^2\right) + \frac{l^2}{r_p^2} dr_p^2,
\end{equation}
with $t_p < 0$. The spacetime singularity is now at $t_p = 0$ as shown in Fig \ref{fig:penrose:poincare:patch}. This special case can be used to understand the no production result in Section \ref{section:open:string} because with the the Poincar\'e coordinates they can be shown to be explicitly adiabatic.

\section{D3 Brane Radiation} \label{section:brane:radiation}
The action of a probe D3 brane in the aforementioned background is
\begin{equation}
	I_{D3} = - T_3\int d^4\xi \sqrt{-\det g_{\mu\nu}\partial_a x^{\mu}\partial_b x^{\nu}} + \int A^{(4)},
\end{equation}
with 
\begin{equation}
	\label{equation:brane:tension}
	T_3 = \frac{1}{\left(2\pi\right)^3 g_s \alpha'^2}
\end{equation}
being the tension of the brane and $A^{(4)}$ the Ramond-Ramond potential of the five-flux background. In the rest of the paper we will work with the Gullstrand-Painlev\'e coordinates unless otherwise noted and drop the subscript of the time coordinate. With this choice of coordinate system, the potential becomes
\begin{equation}
	A^{(4)} = \frac{r^4}{l}dt\wedge \sqrt{\sigma} d^3\sigma.
\end{equation}
With the gauge choice $\xi^a = (t, \vec{\sigma})$ and the motion restricted to the $t$ and $r$ directions for simplicity, the action of the brane becomes
\begin{equation}
	I_{D3} = -T_3 V_s\int dt\frac{r^4}{l}\left[\sqrt{1-\frac{l^4}{r^4}\left(\frac{dr}{dt}+\frac{r}{l}\sqrt{1+\frac{\mu}{r^2}}\right)^2} - 1\right]
\end{equation}

The brane emission can be treated as quantum tunneling in a way very similar to early works on Hawking radiation \cite{Parikh:1999mf, Kraus:1994by}. The tunneling rate can be approximately computed by 
\begin{equation}
	\label{equation:radiation:rate:imaginary:action}
	\Gamma \sim e^{-2 \mathrm{Im} I_{D3}},
\end{equation}
where the D3 brane action is evaluated along a classical path that crosses the horizon with the pole shifted via the standard $i\epsilon$ prescription. In this section we compute this rate using a method based on the Hamilton-Jacobi formalism. In Appendix \ref{appendix:checking:black:hole:radiation} we apply this method in two well studied Hawking radiation calculation and show that our method reproduces the expected results.

The momentum conjugate to $r$ is 
\begin{equation}
	\label{equation:conjugate:momentum}
	p = \frac{\delta I_{D3}}{\delta \dot{r}} = T_3 V_s l^3 \frac{\dot{r} + \frac{r}{l}\sqrt{1 + \frac{\mu}{r^2}}}{\sqrt{1 - \frac{l^4}{r^4}\left(\dot{r} + \frac{r}{l}\sqrt{1 + \frac{\mu}{r^2}}\right)^2}}
\end{equation}
and the probe D3 brane Hamiltonian is
\begin{equation}
	\label{equation:hamiltonian}
	H_{D3} = \dot{r}p - L = \frac{r^2}{l^2}\sqrt{p^2 + T_3^2 V_s^2 l^2 r^4} - \frac{r}{l}p\sqrt{1 + \frac{\mu}{r^2}} - T_3 V_s \frac{r^4}{l}.
\end{equation}
Since the Hamiltonian does not have an explicit time dependence, it is a conserved quantity. Therefore in the Hamilton-Jacobi formalism, Hamilton's principal function can be written as 
\begin{equation}
	S = W(r) - \omega t,
\end{equation}
where $W(r)$ is a function of $r$ that we will soon determine and $\omega$ is the energy of the probe brane. Plugging it into the Hamilton-Jacobi equation we have
\begin{equation}
	\label{equation:hamilton:jacobi}
	H_{D3}(r, p, t) + \frac{\partial S}{\partial t} = 0 ~\Longrightarrow ~H_{D3}\left(r, \frac{\partial W}{\partial r}, t\right) - \omega = 0.
\end{equation}
Solving (\ref{equation:hamilton:jacobi}) then gives
\begin{align}
	p = \frac{\partial W}{\partial r} = \frac{l}{rf(r)}\Big[&\sqrt{1+\frac{\mu}{r^2}}\left(\omega+T_3 V_s \frac{r^4}{l}\right) \nonumber \\%
			&\pm \sqrt{\left(1+\frac{\mu}{r^2}\right)\left(\omega + T_3 V_s \frac{r^4}{l}\right)^2 + f(r)\omega\left(\omega + 2T_3 V_s \frac{r^4}{l}\right)}\Big],
			\label{equation:momentum:in:r}
\end{align}
where ``$-$'' and ``$+$'' signs correspond to in-falling branes and radiated branes, respectively. For the in-falling case, the brane can pass through the horizon classically and there is no pole in $p$ at the horizon crossing, which can be explicitly shown by the following expression of $p$, which is derived from (\ref{equation:momentum:in:r})
\begin{equation}
	\label{equation:momentum:infalling:in:r}
	p = -\frac{\frac{l}{r}\omega\left(\omega + 2T_3 V_s\frac{r^4}{l}\right)}{\sqrt{1+\frac{\mu}{r^2}}\left(\omega + T_3 V_s\frac{r^4}{l}\right) + \sqrt{\left(1 + \frac{\mu}{r^2}\right)\left(\omega + T_3 V_s \frac{r^4}{l}\right)^2 + f(r) \omega \left(\omega + 2T_3 V_s \frac{r^4}{l}\right)}}
\end{equation}

\begin{figure}
	\centering
	\begin{tikzpicture}
		\draw[thick, ->] (-4, 0) -- (-2,0);
		\draw[dashed, -] (-2, 0) -- (2,0);
		\draw[thick, ->] (2,0) -- (4,0);
		\draw[thick, ->] (-2, 0) node[above]{$r_h - \epsilon$} to [out=270, in=180] (0,-2) ;
		\draw[thick] (0, -2) to [out=0, in=270] (2, 0) node[above]{$r_h + \epsilon$};
		\draw[fill=black] (0,0) circle (0.5ex);
		\draw node[above] (0,0){$r_h$};
	\end{tikzpicture}
	\caption{The path for evaluating the imaginary part of the action (\ref{equation:brane:imginary:action}), which comes from going around the pole along the semi-circle.}.
	\label{fig:pole:prescription}
\end{figure}
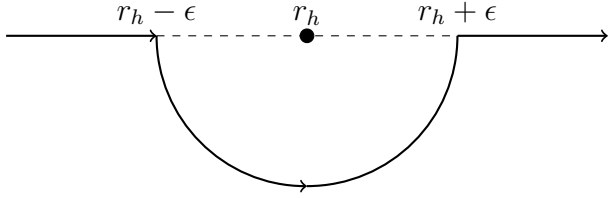

The solution for out-going branes does have a pole at the horizon since there is no classical path escaping the black hole from the inside. However one can still evaluate the action for a path crossing the horizon quantum mechanically using the usual $i\epsilon$ prescription and the imaginary part this procedure generates represents the tunneling amplitude across the horizon as follows. To start with, let us note that the value of an action is the same as Hamilton's principal function when they are evaluated with the same solution. Therefore we can compute the imaginary part of the former by computing the imaginary part of latter over the horizon-crossing region
\begin{equation}
	\label{equation:radiation:imaginary:setup}
	\mathrm{Im} I = \mathrm{Im}\left[ \int_{r_h-\delta}^{r_h + \delta}dr \left(\frac{\partial W}{\partial r}\right) - \omega t\right],
\end{equation}
where $\delta >0$ is an infinitesimal parameter. The integral has a pole at $r_h$, which comes from the $f(r) \simeq f'(r_h)(r-r_h)$ factor in the denominator of (\ref{equation:momentum:in:r}). Properly using the $i\epsilon$ prescription and having the integral contour go around the pole would generate the desired imaginary part. 
To do so, let us temporarily consider back-reaction, so that after one brane tunnels out the black hole radius would shrink, namely $r_h \to r_h - \Delta r_h(\omega)$ and $\Delta r_h(\omega)$ is a monotonically increasing function of $\omega$, the energy of the brane. In the usual $i\epsilon$ prescription, $\omega$ is replaced by $\omega - i\epsilon$, which then puts the pole at
\begin{align}
	r_h - \Delta r_h(\omega - i\epsilon) \simeq r_h -\left[\Delta r_h(\omega) -i\epsilon\frac{d\Delta r_h}{d\omega}\right] \simeq r_h -\Delta r_h + i\epsilon,
\end{align}
where we have used $i\epsilon\frac{d\Delta r_h}{d\omega} \to i\epsilon$ since $\epsilon$ denotes an infinitesimal positive number. The action integral now effectively goes around the pole along a semi-circle, as shown in Fig \ref{fig:pole:prescription}. Defining $r - r_h = \epsilon e^{i\theta}$ where $\theta$ goes from $-\pi$ to 0 and only keeping the leading order contribution (without back-reaction) we have
\begin{align}
	\label{equation:brane:imginary:action}
	\mathrm{Im} I & = \mathrm{Im}\int_{-\pi}^0 d\left(\epsilon e^{i\theta}\right)
				\frac{l}{r_h f'(r_h) \left(\epsilon e^{i\theta}\right)}\left[2\sqrt{1+\frac{\mu}{r_h^2}}\left(\omega + T_3 V_s \frac{r_h^4}{l}\right)\right] \nonumber \\%
				&= \frac{2\pi}{f'(r_h)}\left(\omega + T_3 V_s \frac{r_h^4}{l}\right).
\end{align}
Notice that the the imaginary part comes from the infinitesimal contour integral around the pole (picking up half of the residue), not from an integral of a complex function over a finite region as in instanton computations. This also happens in the calculation of \cite{Parikh:1999mf}.
With this result, the radiation rate is then
\begin{equation}
	\Gamma \sim \exp\left[-\frac{4\pi}{f'(r_h)}\left(\omega + T_3 V_s \frac{r_h^4}{l}\right)\right] = \exp\left[-\beta \left(\omega + T_3 V_s \frac{r_h^4}{l}\right)\right],
\end{equation}
where $\beta = 4\pi/f'(r_h)$ is the inverse temperature. It can also be expressed in terms of number of formation branes using (\ref{equation:brane:tension}) and (\ref{equation:l:in:N}) 
\begin{equation}
	\label{equation:radiation:rate:in:N}
	\Gamma \sim \exp\left(-\frac{V_s \xi^5}{\pi\left(2\xi^2 - 1\right)}N - \frac{2\pi\xi}{2\xi^2 - 1}\left(4\pi g_s N\right)^{\frac{1}{4}}\omega\sqrt{\alpha'}\right),
\end{equation}
with $\xi = r_h / l$. Depending on the energy the radiated brane takes away, we can schematically write down the time it takes to radiate one brane as
\begin{equation}
	\Delta t = t_0 \exp\left(C N^p\right),
\end{equation}
where $C$ and $p$ are two constants and $t_0$ is the fundamental time scale for the vacuum fluctuation of D3 branes. For our purpose the exact value of this constant is not important since we mostly care about the parametric dependence on $N$. The value of the order 1 parameter $p$ depends on the relative size of the first and second terms in (\ref{equation:radiation:rate:in:N}): $p=1$ when the first term dominates and $p > 1$ otherwise. However it needs to be of order 1 since we are considering the regime where there is no back-reaction on the black hole background. In Section \ref{section:open:string} we will see that there is a well-defined open string production regime with small back-reaction. 

From the exponential dependence on $N$ it is clear that the time separation between two branes is very large. One can also roughly estimate the lifetime of such black holes as
\begin{equation}
	t_0 e^{CN^p} + t_0 e^{C(N-1)^p} + \cdots = t_0 e^{CN^p}\left(1 + e^{-pCN^{p-1}} + e^{-2pC N^{p-1}} + \cdots\right) \simeq t_0 e^{CN^p},
\end{equation}
which indicates that the time of radiating one brane is at the same order with the black hole lifetime. This seems to line up with the time line of the advent of the putative firewall according to the AMPS paradox.

\section{Open String Pair Production} \label{section:open:string}
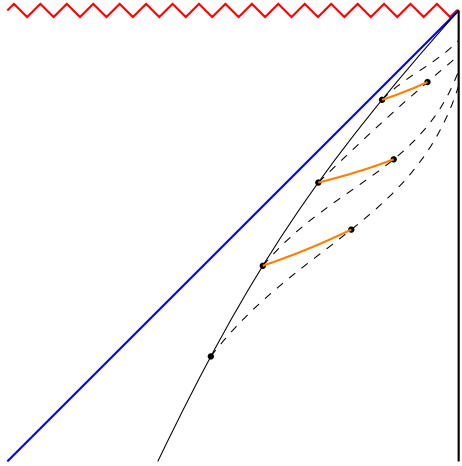
\begin{figure}
	\centering
	\begin{tikzpicture}
		\draw[decorate, decoration=zigzag, red, thick] (-3,0) -- (3,0);
		\draw[blue, thick] (-3, -6) -- (3,0);
		\draw[thick] (3, -6) -- (3, 0);
		\draw (-1, -6) to [bend left=8] coordinate[pos=0.2] (P1) coordinate[pos=0.4] (P2)  coordinate[pos=0.6] (P3) coordinate[pos=0.8] (P4) (3,0);
		
		\fill[black] (P1) circle (1.2pt);
		\fill[black] (P2) circle (1.2pt);
		\fill[black] (P3) circle (1.2pt);
		\fill[black] (P4) circle (1.2pt);
		
		\draw[dashed] (P1) to [bend left=5, in=205] coordinate[pos=0.5] (P5) (3, -1);
		\draw[dashed] (P2) to [bend left=5, in=205] coordinate[pos=0.6] (P6)(3, -0.8);
		\draw[dashed] (P3) to [out=44, in=220] coordinate[pos=0.8] (P7) (3, -0.6);
		\draw[dashed] (P4) to [out=45, in=225] (3, -0.4);
		
		\fill[black] (P5) circle (1.2pt);
		\fill[black] (P6) circle (1.2pt);
		\fill[black] (P7) circle (1.2pt);
		
		\draw[orange, thick] (P2) to [bend right=3] (P5);
		\draw[orange, thick] (P3) to [bend right=3] (P6);
		\draw[orange, thick] (P4) to [bend right=3] (P7);
	\end{tikzpicture}
	\caption{Open string pair production between radiated branes. The branes, whose world-volumes are represented by the dashed lines, start a little outside the horizon and move towards the boundary. Note that for simplicity we have put their starting points on the same constant $r$ curve, represented by the black solid curve. Also its distance to the horizon has been exaggerated for showing the relevant part - they should have started much closer to the horizon. The produced open strings are schematically shown by the orange curves stretched between the dashed curves.}
	\label{fig:open:string:production}
\end{figure}

With the classical dynamics and radiation of D3 probe branes worked out, we can now compute the level of pair production of open strings stretched betweeen two consecutively radiated branes, as schematically shown in Fig \ref{fig:open:string:production}. In the following we will use a real time estimation method based on the non-adiabaticity figure of merit $\omega^2/\dot{\omega}$ as discussed in \cite{Silverstein:2014yza, Puhm:2016sxj}, where $\omega$ is the frequency of open strings and dot means derivative with respect to the time coordinate under use. More discussion about this method is given below after (\ref{equation:string:action:ansatz:full}).

To begin with, let us write down the following symmetric ansatz for open strings stretched between two branes separated by time $\Delta t$
\begin{equation}
	\label{equation:open:string:ansatz}
	t = t(\tau) + \Delta t \frac{\sigma}{\pi}, \quad r= r(\tau), \quad X_{\perp} = b_S\frac{\sigma}{\pi},
\end{equation}
where $X_{\perp}$ represents the $S^5$ directions and $b_S$ the impact parameter of the branes in those directions. The trajectory of the first brane is encoded in the functions $t(\tau)$ and $r(\tau)$. Ansatz (\ref{equation:open:string:ansatz}) can explicitly solve the Dirichlet boundary conditions in the $t, r$ and $S^5$ directions as derived from the boundary terms of the worldsheet action. The strings are allowed to move freely in the $\Sigma$ directions around which the branes wrap and in principle we need to impose Neumann boundary conditions for them. In ansatz (\ref{equation:open:string:ansatz}) we ignore those directions for simplicity, assuming that the strings stay at fixed location in those directions.

Ansatz (\ref{equation:open:string:ansatz}) can be used to compute the Nambu-Goto action
\begin{equation}
	S = -\frac{1}{\alpha'}\int d\tau d\sigma \sqrt{-\det \gamma_{ab}},
\end{equation}
with $\gamma_{ab} = g_{\mu\nu}\partial_a X^{\mu}\partial_b X^{\nu}$ being the induced metric on the open string worldsheet. Using Gullstrand-Painlev\'e coordinates we have
\begin{align}
	\gamma_{\tau\tau} &= -\left(\frac{r^2}{l^2} - 1-\frac{\mu}{r^2}\right)t'^2 + 2\frac{l}{r}\sqrt{1+\frac{\mu}{r^2}}t' r' + \frac{l^2}{r^2}r'^2 \\%
	\gamma_{\tau\sigma} &= -\left(\frac{r^2}{l^2} - 1 - \frac{\mu}{r^2}\right)t'\frac{\Delta t}{\pi} + \frac{l}{r}\sqrt{1 + \frac{\mu}{r^2}}r'\frac{\Delta t}{\pi} \\%
	\gamma_{\sigma\sigma} &= -\left(\frac{r^2}{l^2} - 1 - \frac{\mu}{r^2}\right)\left(\frac{\Delta t}{\pi}\right)^2 + \left(\frac{b_S}{\pi}\right)^2,
\end{align}
where prime $'$ means the derivative with respective to worldsheet coordinate $\tau$. The open string action then evaluates to 
\begin{equation}
	\label{equation:open:string:action:GP:time}
	S = -\frac{1}{\alpha'}\int dt \sqrt{\Delta t^2 \left(\frac{dr}{dt}\right)^2 + b_S^2 \frac{r^2}{l^2}\left[1 - \frac{l^4}{r^2}\left(\frac{dr}{dt} + \frac{r}{l}\sqrt{1+\frac{\mu}{r^2}}\right)^2\right]}
\end{equation}

To estimate the level of non-adiabaticity we need to pick a time coordinate. The Gullstrand-Painlev\'e coordinate $t$ is not a very good choice since it is conformal to the proper time of an auxiliary observer falling into the black hole from infinity and thus might generate unnecessary non-adiabaticity due to the relative boost between radiated branes and the auxiliary brane. To eliminate this problem we use the proper time of the radiate branes, as determined by
\begin{equation}
	d\tau^2 = \frac{r^2}{l^2} dt^2 \left[1 - \frac{l^4}{r^4}\left(\dot{r} + \frac{r}{l}\sqrt{1 + \frac{\mu}{r^2}}\right)^2\right]
\end{equation}
with $\dot{r}$ satisfying the equation of motion of the radiated branes\footnote{Note that when deriving the worldsheet action we use $\tau$ to denote the worldsheet time coordinate but here $\tau$ is the proper time of the radiated branes. In the rest of this section when we analyze the level of non-adiabaticity $\tau$ always means the latter.}. With this time coordinate (\ref{equation:open:string:action:GP:time}) becomes 
\begin{equation}
	S = -\frac{1}{\alpha'}\int d\tau \sqrt{b_S^2 + \Delta t^2\left(\frac{dr}{d\tau}\right)^2}.
\end{equation}
The time derivative of $r$ can be better computed in the following way. Define $X$ as
\begin{equation}
	X = \dot{r} + \frac{r}{l}\sqrt{1+\frac{\mu}{r^2}}
\end{equation}
then from the definition of $p$ (\ref{equation:conjugate:momentum}) and the out-going case of (\ref{equation:momentum:in:r}) we have
\begin{equation}
	\frac{X}{\sqrt{1-\frac{l^4}{r^4}X^2}} = \frac{p}{T_3 V_s l^3} = \frac{1}{f(r)}\left[\sqrt{1+\frac{\mu}{r^2}}\left(\frac{r^3}{l^3} + C\frac{l}{r}\right) + 
		\sqrt{\left(1 + \frac{\mu}{r^2}\right)\frac{r^6}{l^6} + C\frac{r^4}{l^4}\left(2 + C\frac{l^4}{r^4}\right)}\right],
\end{equation}
with the constant $C = \omega/ (T_3 V_s l^3)$ playing a similar role as the constant $C$ in previous works \cite{Silverstein:2014yza, Puhm:2016sxj}. It quantifies how much energy the brane carries away in the units of its ``rest mass'' and how boosted the branes are at the horizon. We will discuss various open string production scenarios based on the size of this parameter.

In the meantime we can express $dr/d\tau$ as
\begin{equation}
	\frac{dr}{d\tau} = \frac{X-\frac{r}{l}\sqrt{1+\frac{\mu}{r^2}}}{\frac{r}{l}\sqrt{1-\frac{l^4}{r^4}X^2}} = \frac{l}{r}\frac{p}{T_3 V_s l^3} - \frac{l^2}{r^2}\sqrt{1 + \frac{\mu}{r^2}}\sqrt{\left(\frac{p}{T_3 V_s l^3}\right)^2 + \frac{r^4}{l^4}}.
\end{equation}
Using (\ref{equation:hamiltonian}) and $H_{D3} = \omega$ it can be further reduced to 
\begin{equation}
	\frac{dr}{d\tau} = \sqrt{1 + \frac{\mu}{r^2} + C\frac{l^3}{r^3}\left(2\frac{r}{l} + C\frac{l^3}{r^3}\right)},
\end{equation}
which can be plugged into the worldsheet action and generates
\begin{equation}
	\label{equation:string:action:ansatz:full}
	S = -\frac{1}{\alpha'}\int d\tau \sqrt{b_S^2 + n\alpha' + \Delta t^2\left[1 + \frac{\mu}{r^2} + C\frac{l^3}{r^3}\left(2\frac{r}{l} + C\frac{l^3}{r^3}\right)\right]},
\end{equation}
where oscillator level $n$ has been included as in \cite{Silverstein:2014yza, Puhm:2016sxj, McAllister:2004gd, Bachlechner:2013fja, Bachas:1995kx, Bachas:1992bh}. 

In principle any quantum amplitude should be defined by the path integral of $e^{iS}$ and the production amplitude is given by the imaginary part of the action evaluated at the proper saddle points. By symmetry a saddle point of ansatz (\ref{equation:open:string:ansatz}) is a saddle point of the full theory. Therefore the complete calculation of open string production rate should be to first determine the complex roots of the frequency in (\ref{equation:string:action:ansatz:full})
\begin{align}
	\omega_{\tau} = -\frac{1}{\alpha'}\sqrt{b_S^2 + n\alpha' + \Delta t^2\left[1 + \frac{\mu}{r^2} + C\frac{l^3}{r^3}\left(2\frac{r}{l} + C\frac{l^3}{r^3}\right)\right]}
\end{align}
and then compute the complex action using contours around those roots. More details about this method can be found in \cite{Silverstein:2014yza}. However this full calculation seems difficult to perform for our case, due to the complicated dependence of $\omega_{\tau}$ on $r$ and $\tau$. To proceed we realize that the production stems from the non-adiabaticity of $\omega_{\tau}$ near the horizon. It has been argued in previous works \cite{Silverstein:2014yza, Puhm:2016sxj} that
\begin{equation}
	\frac{d\omega_{\tau}/d\tau}{\omega_{\tau}^2}\Big|_{r=r_h} = \frac{\Delta t^2 \alpha'}{r_h} \frac{\left(\frac{r_h^2}{l^2} - 1 + 2C\frac{l^2}{r_h^2} + 3C^2\frac{l^6}{r_h^6}\right)\left(\frac{r_h}{l} + C\frac{l^3}{r_h^3}\right)}{\left[b_S^2 + n\alpha' + \Delta t^2\left(\frac{r_h}{l} + C\frac{l^3}{r_h^3}\right)^2\right]^{\frac{3}{2}}}.
\end{equation}
is a good figure of merit for non-adiabaticity and we can estimate the production density of strings at oscillator level $n$ as 
\begin{align}
	\exp\left(-\frac{\omega_{\tau}^2}{d\omega_{\tau}/d\tau}\right).
\end{align}
This gives the contribution from a single string saddle point at oscillator level $n$. To get the full production density, we fold in the Hagedorn density at level $n$
\begin{equation}
	\exp\left(\sqrt{8\pi^2 n}\right)
\end{equation}
and sum over all the oscillator levels. The resulting full density is
\begin{equation}
	\rho_{\mathrm{tot}} \sim \sum_n e^{K(n)}
\end{equation}
with
\begin{equation}
	\label{equation:string:production:in:n}
	K(n) = \sqrt{8\pi^2 n} - \frac{r_h}{\Delta t^2 \alpha'}\frac{\left[n\alpha' + \Delta t^2 \left(\frac{r_h}{l} + C\frac{l^3}{r_h^3}\right)^2\right]^{\frac{3}{2}}}{\left(\frac{r_h^2}{l^2} - 1 +2C\frac{l^2}{r_h^2} + 3C^2\frac{l^6}{r_h^6}\right)\left(\frac{r_h}{l} + C\frac{l^3}{r_h^3}\right)},
\end{equation}
where we have set $b_S = 0$ for simplicity. 

In Subsection \ref{subsection:no:production:open:string} we analyze the regime where there is no open string production. This is the regime where $C$ is parametrically small. We also study this no production result for the $\mu = 0$ case using Poincar\'e coordinates, where the non-adiabaticity can be demonstrated manifestly. In Subsection \ref{subsection:production:open:string} we analyze the regime where there is enhanced production.

\subsection{No production with $C \ll 1$}\label{subsection:no:production:open:string}
The peak value of $K$ can be determined by taking its derivative with respect to $n$. In the case of $C \ll 1$ the derivation can be simplified by $n\alpha' \ll \Delta t^2 \left(\frac{r_h}{l} + C\frac{l^3}{r_h^3}\right)^2$ (the exact condition for this approximation is (\ref{equation:condition:no:production:reduced})), which we verify in a moment using the peak location $n$ in this case. Setting $K'(n) = 0$ yields 
\begin{equation}
	\sqrt{n} \simeq \frac{\sqrt{8\pi^2}\left(\frac{r_h^2}{l^2} - 1 + 2C\frac{l^2}{r_h^2} + 3C^2\frac{l^6}{r_h^6}\right)\Delta t}{3r_h}.
\end{equation}
To make this solution consistent with the assumption $n\alpha' \ll \Delta t^2 \left(\frac{r_h}{l} + C\frac{l^3}{r_h^3}\right)^2$, it must be much smaller than
\begin{equation}
	\frac{\Delta t\left(\frac{r_h}{l} + C\frac{l^3}{r_h^3}\right)}{\sqrt{\alpha'}},
\end{equation}
which can be expressed by the following inequality
\begin{equation}
	\label{equation:condition:no:production}
	\left(\frac{r_h^2}{l^2} + C\frac{l^2}{r_h^2} - 1\right)\frac{\sqrt{\alpha'}}{l} + 3C^2\frac{l^6}{r_h^6}\frac{\sqrt{\alpha'}}{l} \ll \frac{r_h^2}{l^2} + C\frac{l^2}{r_h^2}.
\end{equation}
Since $r_h/l \ge 1/\sqrt{2}$ and $\sqrt{\alpha'} \ll l$, a sufficient condition for (\ref{equation:condition:no:production}) is 
\begin{equation}
	\label{equation:condition:no:production:reduced}
	C\frac{l^3}{r_h^3}\ll \frac{r_h}{l}\sqrt{\frac{l}{\sqrt{\alpha'}}}.
\end{equation}
In this regime the peak value of $K(n)$ is 
\begin{align}
	&\frac{\Delta t}{3r_h\left(\frac{r_h^2}{l^2} - 1 + 2C\frac{l^2}{r_h^2} + 3C^2\frac{l^6}{r_h^6}\right)}\left[\sqrt{8\pi^2}\left(\frac{r_h^2}{l^2} - 1+2C\frac{l^2}{r_h^2} + 3C^2\frac{l^6}{r_h^6}\right) - \frac{\sqrt{3}l}{\sqrt{\alpha'}}\left(\frac{r_h^2}{l^2}+C\frac{l^2}{r_h^2}\right)\right] \nonumber \\%
		&\times \left[\sqrt{8\pi^2}\left(\frac{r_h^2}{l^2} - 1+2C\frac{l^2}{r_h^2} + 3C^2\frac{l^6}{r_h^6}\right) + \frac{\sqrt{3}l}{\sqrt{\alpha'}}\left(\frac{r_h^2}{l^2}+C\frac{l^2}{r_h^2}\right)\right]
\end{align}
The difference term is negative due to condition (\ref{equation:condition:no:production}). Since $\Delta t$ is exponential in $N$, $K(n)$ is much smaller than zero in this regime and therefore open string production is highly suppressed.

To understand this no production result from a different perspective, let us analyze this with the Poincar\'e coordinates. As mentioned earlier, this is possible because the black hole with $\mu = 0$ (or $r_h = l$) is isometric to pure AdS$_5$. The equation of motion of D3 probe branes in this background can be solved exactly for the case of $C = 0$ where it simplifies to 
\begin{equation}
	\frac{X}{\sqrt{1-\frac{l^4}{r^4}X^2}} = \frac{1}{f(r)}\left(\frac{r^3}{l^3} \pm \frac{r^3}{l^3}\right).
\end{equation}
The in-falling solution is just $X=0$ or $r = r_0 e^{-(t-t_0)/l}$. As discussed in \cite{Horowitz:2009wm} such solutions correspond to stationary branes with $r_p = const.$ and therefore the in-falling case is manifestly adiabatic.

For the case of radiated branes, the equation of motion reduces to
\begin{equation}
	\frac{dr}{dt} = \frac{r}{l}\frac{r^2 - l^2}{r^2 + l^2}, \qquad r \ge l
\end{equation}
whose solution is
\begin{equation}
	\label{equation:solution:outgoing:no:production}
	\frac{\left(r^2 - l^2\right)r_i}{\left(r_i^2 - l^2\right)r} = e^{\frac{t-t_i}{l}},
\end{equation}
where $r_i$ and $t_i$ denote the location and time the branes materialize, respectively.
It is clear that $r(t)$ grows exponentially with $t$ and (\ref{equation:solution:outgoing:no:production}) is approximately
\begin{equation}
	r \simeq \frac{r_i^2 - l^2}{r_i} e^{\frac{t-t_i}{l}}.
\end{equation}
Switching to the Poincar\'e coordinates using (\ref{equation:GP:2:Poincare}) we have the trajectory for the first brane as 
\begin{equation}
	r_p^{(1)} \simeq \frac{r_i^2 - l^2}{r_i}e^{-\frac{t_i}{l}}\left(\frac{l}{t_p}\right)^2.
\end{equation}
For the second brane it is $r_p^{(2)} \simeq e^{-\Delta t/l}r_p^{(1)}$. The proper distance between the two branes is
\begin{equation}
	\Delta r_p = \int_{r_p^{(2)}}^{r_p^{(1)}} \frac{l}{r_p}dr_p = l\log\frac{r_p^{(1)}}{r_p^{(2)}} = \Delta t,
\end{equation}
which is still a constant. Therefore for the case of radiated branes with $C=0$ and $\mu = 0$ it is still manifestly adiabatic, agreeing with the estimation using Gullstrand-Painlev\'e coordinates.

\subsection{Production regime} \label{subsection:production:open:string}
The peak value of $K(n)$ can also be evaluated exactly. By setting $K'(n) = 0$ we have the peak location at
\begin{equation}
	\label{equation:string:production:peak:location}
	n = \frac{-A^2 + A^2\sqrt{1 + \frac{32\pi^2}{9}\frac{B^2}{A^2}\frac{\alpha'}{r_h^2}}}{2}\frac{\Delta t^2}{\alpha'},
\end{equation}
where for notational simplicity we have defined the following constants
\begin{equation}
	\label{equation:A:B:definition}
	A = \frac{r_h}{l} + C\frac{l^3}{r_h^3}, \quad B = \frac{r_h^2}{l^2} - 1+2C\frac{l^3}{r_h^3}\frac{r_h}{l} + 3C^2\frac{l^6}{r_h^6}.
\end{equation}
Plugging (\ref{equation:string:production:peak:location}) into (\ref{equation:string:production:in:n}), the peak value of $K(n)$ is
\begin{equation}
	\label{equation:K:max}
	K_{\mathrm{max}} = \frac{\Delta t}{\left(1+\sqrt{1 + \frac{32\pi^2}{9}\frac{B^2}{A^2}\frac{\alpha'}{r_h^2}}\right)^{\frac{1}{2}}}\frac{r_h}{\alpha'}\frac{A^2}{B}
	\left[\frac{16\sqrt{2}\pi^2}{9}\frac{B^2}{A^2}\frac{\alpha'}{r_h^2} - \frac{\sqrt{2}}{2}\left(1+\sqrt{1+\frac{32\pi^2}{9}\frac{B^2}{A^2}\frac{\alpha'}{r_h^2}}\right)\right].
\end{equation}
As discussed earlier, $\Delta t$ is an exponentially large parameter, therefore the production effect will be enhanced once the difference factor in the square bracket becomes positive. This happens roughly near the region where
\begin{equation}
	\label{equation:condition:production}
	\frac{B^2}{A^2}\frac{\alpha'}{r_h^2} \gtrsim 1.
\end{equation}
Looking at the definition of $A$ and $B$ (\ref{equation:A:B:definition}) it is clear that we need $C\frac{l^3}{r_h^3}\gg \frac{r_h}{l}$ to satisfy (\ref{equation:condition:production}) since otherwise $B/A \simeq r_h^2/l^2$ and $(B^2/A^2)(\alpha'/r_h^2) \simeq \alpha'/l^2 \ll 1$. On the other hand with the right condition ($C\frac{l^3}{r_h^3}\gg \frac{r_h}{l}$) we have
\begin{equation}
	\frac{B}{A} \simeq 3C\frac{l^3}{r_h^3}.
\end{equation}
This simplification reduces the difference factor in (\ref{equation:K:max}) to
\begin{equation}
	16\sqrt{2}\pi^2 C^2\frac{l^6}{r_h^6}\frac{\alpha'}{l^2}\frac{l^2}{r_h^2} - \frac{\sqrt{2}}{2}\left(1 + \sqrt{1 + 32\pi^2 C^2\frac{l^6}{r_h^6}\frac{\alpha'}{l^2}\frac{l^2}{r_h^2}}\right),
\end{equation}
where we have factored $\sqrt{\alpha'}/r_h$ into $(\sqrt{\alpha'}/l)(l/r_h)$ to show the different independent parameters we use. This difference is positive in the regime where
\begin{equation}
	\label{equation:condition:production:reduced}
	C\frac{l^3}{r_h^3}\frac{\sqrt{\alpha'}}{l}\frac{l}{r_h} \ge 1 \Longrightarrow C\frac{l^3}{r_h^3} \ge \frac{l}{\sqrt{\alpha'}}\frac{r_h}{l},
\end{equation}
which is the consistent with the condition $C\frac{l^3}{r_h^3}\gg \frac{r_h}{l}$. It is intriguing to notice that this condition does not directly rely on the time separation between two branes $\Delta t$. This is also observed in earlier works \cite{Silverstein:2014yza, Puhm:2016sxj}.

\begin{figure}
	\centering
	\begin{subfigure}[b]{0.45\textwidth}
		\includegraphics[width=\textwidth]{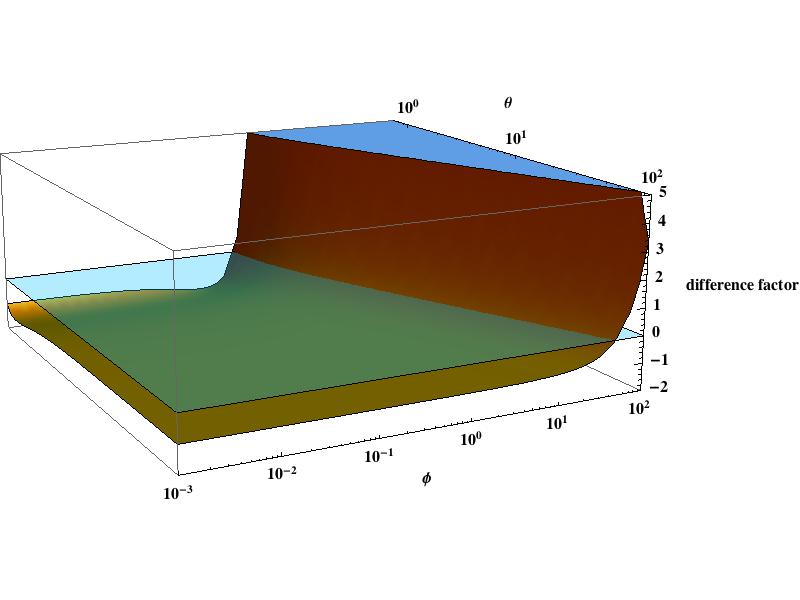}
		\caption{$g_s = 0.1, N = 1000$}
	\end{subfigure}
	\begin{subfigure}[b]{0.45\textwidth}
		\includegraphics[width=\textwidth]{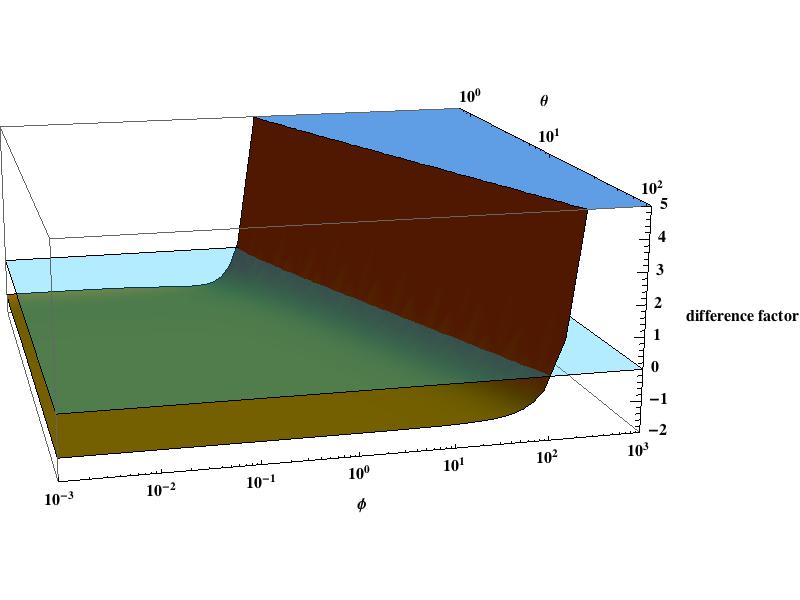}
		\caption{$g_s = 0.1, N = 10000$}
	\end{subfigure}
	\begin{subfigure}[b]{0.45\textwidth}
		\includegraphics[width=\textwidth]{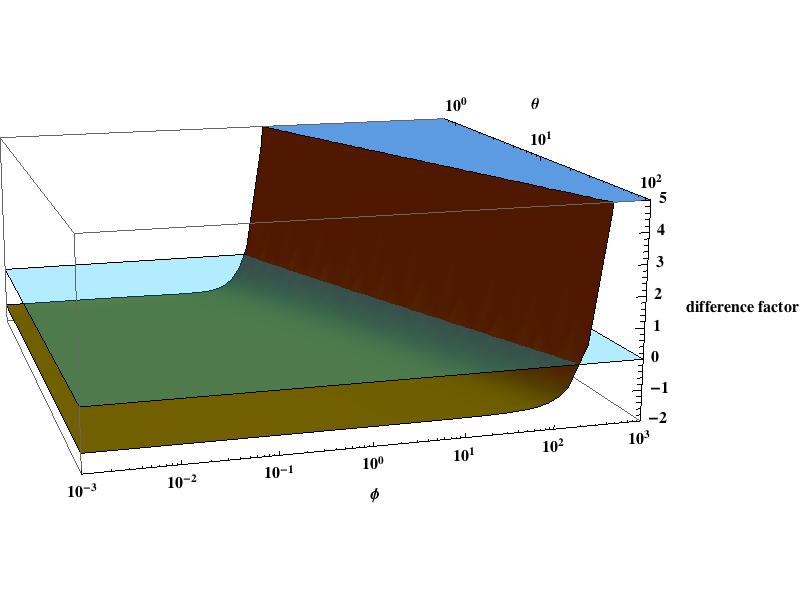}
		\caption{$g_s = 0.1, N = 100000$}
	\end{subfigure}
	\begin{subfigure}[b]{0.45\textwidth}
		\includegraphics[width=\textwidth]{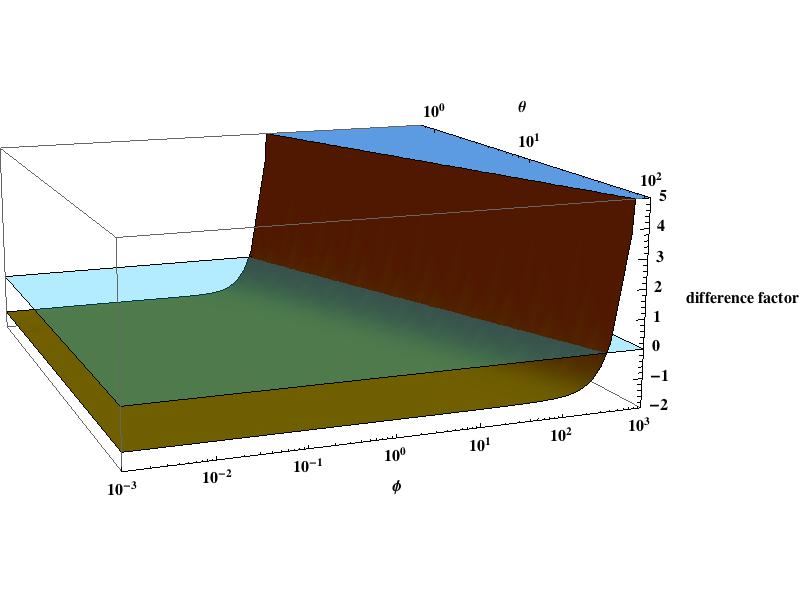}
		\caption{$g_s = 0.1, N = 1000000$}
	\end{subfigure}
	\caption{Plots of the difference factor in (\ref{equation:K:max}). The curved surface represents the difference factor as a function of $\phi$ and $\theta$, which are defined in (\ref{equation:theta:phi:definition}). The horizontal surface in cyan is where the difference is zero. Its intersection with the curved surface represents the onset of non-adiabatic production. This indeed confirms the condition (\ref{equation:condition:production:reduced:in:theta:phi}) derived in main text where the onset roughly follows $\phi \simeq \left(4\pi g_s N\right)^{\frac{1}{4}}\theta$ }
	\label{fig:K:max:difference}
\end{figure}

The condition for open string production (\ref{equation:condition:production:reduced}) can also be checked numerically. In Fig \ref{fig:K:max:difference} we plot the value of the difference factor of (\ref{equation:K:max}) in terms of the following two independent variables
\begin{equation}
	\label{equation:theta:phi:definition}
	\theta = \frac{r_h}{l}, \qquad \phi = C\frac{l^3}{r_h^3}.
\end{equation}
The other two parameters are the string couple $g_s$ and number of formation branes $N$ which only appear as their product $g_s N$. We show four cases depending on the value of $g_s N$. In terms of these parameters the production condition (\ref{equation:condition:production:reduced}) becomes 
\begin{equation}
	\label{equation:condition:production:reduced:in:theta:phi}
	\phi \ge \left(4\pi g_s N\right)^{\frac{1}{4}}\theta.
\end{equation}
This roughly agrees with the result shown in Fig \ref{fig:K:max:difference}.

We also need to check that the production condition (\ref{equation:condition:production:reduced}) is consistent with the condition of no back-reaction. In terms of $\omega$ the former is
\begin{equation}
	\label{equation:condition:production:in:omega}
	\omega \ge T_3 V_s l^3 \cdot \frac{r_h^3}{l^3}\cdot\frac{r_h}{\sqrt{\alpha'}} = \frac{V_s}{2\pi^2}\frac{N}{\sqrt{\alpha'}} \frac{r_h^4}{l^4}.
\end{equation}
Compared with the mass of the black hole (\ref{equation:black:hole:mass}), which is of order $N^{7/4}/g_s^{1/4}$, the right hand side of (\ref{equation:condition:production:in:omega}) is parametrically smaller than the black hole mass, therefore there is a large parameter window where the production condition and small back-reaction condition are consistent.

\section{Summary And Discussion} \label{section:summary}
In this paper we computed the radiation rate of D3 branes with energy $\omega$ from hyperbolic black holes. It is found that the time it takes to radiate such branes is at the same order of the black hole lifetime (at least before back-reaction becomes important). Non-adiabatic open string production effects were studied using a real time estimation method developed in \cite{Silverstein:2014yza}. We demonstrated that the frequency of open strings stretched between two consecutively radiated branes can become highly non-adiabatic when the radiated branes are highly boosted. Consistency checks are made so that there is no non-adiabatic production in well-known low energy regimes and brane radiation does not lead to back-reaction. It is also clear from our calculation that our result does not rely on the geometry or topology of the hyperbolic horizon as long as it is compact so that the branes can wrap around. Therefore the result in this paper can be naturally carried over to AdS black holes with spherical horizons.

Compared with similar open string production thought experiments in previous works, our scenario does not involve interactions other than the one in the near horizon region, no matter what the kinematic parameters are. In cases where the D branes were dropped in to the black hole from outside, there might be early collisions along their world-lines that could obscure the non-adiabatic production near the horizon \cite{Silverstein:2014yza, Puhm:2016sxj}. The non-adiabatic effects near singularity could also add to the unwanted production. In our cases, however, open string production can only happen due to the non-adiabaticity near the horizon, which makes our result cleaner.


The fact that the open string production events can happen spontaneously in the near horizon region and the time for it to occur is roughly of the order of the black hole lifetime makes our result intriguing in the context of the AMPS paradox. It seems to provide an interesting dynamical scenario for the putative firewall to arise, even within this simple model of D branes and open strings. To make the result more convincing, it is worth studying the open string production rate using a more direct calculation, such as the first quantizated method developed in \cite{Silverstein:2014yza}. This method is not the most ideal method in that work since the world-lines of the in-falling D0 particles involve a large spacetime region from the boundary to the singularity and it is hard to interpret the production result as to where exactly it happens. In our scenario, however, the branes's world-line in the $t, r$ directions are always close to the horizon and any production result using the first quantized method can be attributed to the non-adiabatic effects in the near horizon region.

Another method worth exploring is to compute the amount of open string production in ways similar to \cite{Bachas:1995kx, Bachas:1992bh}, where the density of the produced open strings $\Gamma$ is computed from the one-loop density of the vacuum $\mathcal{F}$ as
\begin{equation}
	\Gamma = -2\mathrm{Im}\mathcal{F}.
\end{equation}
To do this one would first need to be able to solve the string theory in the corresponding spacetime background. For our case it seems to be very hard to do so in the AdS$_5$ black hole background. However exact string theory solutions exist in AdS$_3$. In \cite{Maldacena:2000hw} the authors studied the $SL(2, R)$ WZW model describing string theory on the AdS$_3\times \mathcal{M}$ background where $\mathcal{M}$ is compact. The algebra and spectrum are worked out in that work. Since pure AdS$_3$ is isometric to the BTZ black hole, one should be able to use that solution to study the open string production effect near BTZ black holes, with modifications coming from the identification in the transverse direction of AdS$_3$. In many ways the study of open string production near topological black holes in different spacetime dimensions are similar and the rigorous investigation in AdS$_3$ could shed light on the general properties of such phenomenon.

The production result in this paper depends on the radiated branes being highly boosted. Despite that this requirement does not lead to strong back-reaction, the chance of radiating such branes is still much smaller than radiating branes with $\omega = 0$ ($e^{-CN^{5/4}}$ versus $e^{-CN}$). It is possible that the black hole has already radiated many low energy branes before radiating branes boosted enough to produce open strings. To fully understand the open string production effects one would need to study the general case where the two branes have different $\omega$'s. We leave these interesting directions to future work.


\section*{Acknowledgments}
I would like to thank Eva Silverstein for extensive discussion and useful comments on a draft. I am also grateful to Xi Dong and Gonzalo Torroba for helpful discussion which I learned a lot from. I am supported in part by the National Science Foundation under grant PHY-0756174 and NSF PHY11-25915 and by the Department of Energy under contract DE-AC03-76SF00515.

\begin{appendices}
\section{Checking Method For Computing Black Hole Radiation}\label{appendix:checking:black:hole:radiation}
In this section we apply our method of computing black hole radiation rate to Hawking radiation in some well known black hole backgrounds and verify that our method produces the correct results. To be more general, let us consider neutral particle radiation from a generic black hole with the following metric
\begin{equation}
	ds^2 = -f(r)dt^2 + \frac{dr^2}{f(r)} + r^2 d\Omega^2,
\end{equation}
with the asymptotic geometry as $r\to \infty$ being $f(r) \to h(r)$. It is clear that this implies $f(r)$ has one or more roots while $h(r)$ has no root. Some examples are Minkowski black hole with $h(r) =1$ and AdS black holes with $h(r) = r^2/l^2$ where $l$ is the radius of AdS. Redefine the time slice using
\begin{equation}
	dt \to dt - \frac{1}{f(r)}\sqrt{1-\frac{f(r)}{h(r)}},
\end{equation}
one gets the Gullstrand-Painlev\'e metric for this black hole
\begin{equation}
	ds^2 = -h(r) dt^2 + \frac{1}{h(r)}\left(dr + h(r)\sqrt{1-\frac{f(r)}{h(r)}}dt\right)^2 + r^2 d\Omega^2.
\end{equation}

The world-line action for a neutral particle in this background is
\begin{equation}
	I = -m\int d\lambda \sqrt{-g_{\mu\nu}\partial_{\lambda}x^{\mu}\partial_{\lambda}x^{\nu}}.
\end{equation}
Taking the gauge with $\lambda$ set to $t$, the action for symmetrical solutions becomes 
\begin{equation}
	I = -m\int dt\sqrt{h(r) - \frac{1}{h(r)}\left(\dot{r} + h(r)\sqrt{1 - \frac{f(r)}{h(r)}}\right)^2},
\end{equation}
where dot means the derivative with respect to $t$. The momentum conjugate to $r$ and the Hamiltonian can be derived from it and the result is
\begin{align}
	\label{equation	:particle:momemtum}
	p &= \frac{m}{h(r)}\frac{\dot{r} + h(r)\sqrt{1-\frac{f}{h}}}{\sqrt{h - \frac{1}{h}\left(\dot{r}+h\sqrt{1-\frac{f}{h}}\right)^2}} \\%
	\label{equation:particle:hamiltonian}
	H &= \sqrt{p^2 h(r)^2 + m^2 h(r)} - h(r)p\sqrt{1-\frac{f(r)}{h(r)}}.
\end{align}
Similar to the calculation in Section \ref{section:brane:radiation}, Hamilton's principle function can be written as 
\begin{equation}
	S = W(r) - \omega t,
\end{equation}
with $\omega$ being the conserved energy of the particle, then Hamilton-Jacobi equation gives
\begin{equation}
	\sqrt{\left(\frac{\partial W}{\partial r}\right)^2 h(r)^2 + m^2 h(r)} - h(r)\frac{\partial W}{\partial r}\sqrt{1-\frac{f(r)}{h(r)}} = \omega.
\end{equation}
Its solutions are
\begin{equation}
	\frac{\partial W}{\partial	r} = \frac{\omega\sqrt{1-\frac{f}{h}}\pm\sqrt{\omega^2\left(1-\frac{f}{h}\right) + \frac{f}{h}\left(\omega^2 - m^2 h\right)}}{f(r)},
\end{equation}
We are interested in the one with the ``+'' sign which corresponds to out-going particles.

As discussed in the main text, the radiation rate is related to the imaginary part of the action by (\ref{equation:radiation:rate:imaginary:action}), which can be evaluated using the standard $i\epsilon$ prescription. Using (\ref{equation:radiation:imaginary:setup}) we have the result
\begin{equation}
	\mathrm{Im} I = \frac{2\pi\omega}{f'(r_h)},
\end{equation}
where prime denotes the derivative with respect to $r$. Therefore we have the following general result for Hawking radiation
\begin{equation}
	\label{equation:hawking:radiation:general}
	\Gamma \sim \exp\left(-\frac{4\pi\omega}{f'(r_h)}\right) = e^{-\beta\omega},
\end{equation}
with $\beta = 4\pi/f'(r_h)$ being the inverse temperature.

For Schwarzschild black holes with mass $M$ in Minkowski space, the blackening factor is
\begin{equation}
	f(r) = 1 - \frac{2M}{r},
\end{equation}
for which the horizon is at $r_h = 2M$. One can then readily verify that our result (\ref{equation:hawking:radiation:general}) reproduces the well known result
\begin{equation}
	\Gamma \sim e^{-8\pi M\omega}.
\end{equation}
Another example is Reissner-Nordstr\"om black hole with 
\begin{equation}
	f(r) = 1 - \frac{2M}{r} + \frac{Q^2}{r^2}.
\end{equation}
The horizon $r_h$ satisfies
\begin{equation}
	\frac{1}{r_h} = \frac{M - \sqrt{M^2 - Q^2}}{Q^2}.
\end{equation}
This can be used to compute $f'(r_h)$, which can be shown to be 
\begin{equation}
	f'(r_h) = \frac{2\sqrt{M^2 - Q^2}}{\left(M + \sqrt{M^2 - Q^2}\right)^2}.
\end{equation}
One can again verify that our result (\ref{equation:hawking:radiation:general}) reproduces the well known result
\begin{equation}
	\Gamma \sim \exp\left(-\frac{2\pi\omega \left(M + \sqrt{M^2 - Q^2}\right)^2}{\sqrt{M^2 - Q^2}}\right).
\end{equation}

\end{appendices}


\begin{thebibliography}{9}

\bibitem{Silverstein:2014yza} 
  E.~Silverstein,
  ``Backdraft: String Creation in an Old Schwarzschild Black Hole,''
  arXiv:1402.1486 [hep-th].
  
\bibitem{Puhm:2016sxj} 
  A.~Puhm, F.~Rojas and T.~Ugajin,
  ``(Non-adiabatic) string creation on nice slices in Schwarzschild black holes,''
  JHEP {\bf 1704}, 156 (2017)
  doi:10.1007/JHEP04(2017)156
  [arXiv:1609.09510 [hep-th]].
  
\bibitem{Dodelson:2015toa} 
  M.~Dodelson and E.~Silverstein,
  ``String-theoretic breakdown of effective field theory near black hole horizons,''
  arXiv:1504.05536 [hep-th].


\bibitem{Dodelson:2017hyu} 
  M.~Dodelson and E.~Silverstein,
  ``Long-Range Nonlocality in Six-Point String Scattering: simulation of black hole infallers,''
  arXiv:1703.10147 [hep-th].
  
\bibitem{Dodelson:2017emn} 
  M.~Dodelson, E.~Silverstein and G.~Torroba,
  ``Varying dilaton as a tracer of classical string interactions,''
  arXiv:1704.02625 [hep-th].
  
\bibitem{McAllister:2004gd} 
  L.~McAllister and I.~Mitra,
  ``Relativistic D-brane scattering is extremely inelastic,''
  JHEP {\bf 0502}, 019 (2005)
  doi:10.1088/1126-6708/2005/02/019
  [hep-th/0408085].

\bibitem{Bachlechner:2013fja} 
  T.~C.~Bachlechner and L.~McAllister,
  ``D-brane Bremsstrahlung,''
  JHEP {\bf 1310}, 022 (2013)
  doi:10.1007/JHEP10(2013)022
  [arXiv:1306.0003 [hep-th]].
  
\bibitem{Amati:1987uf} 
  D.~Amati, M.~Ciafaloni and G.~Veneziano,
  ``Classical and Quantum Gravity Effects from Planckian Energy Superstring Collisions,''
  Int.\ J.\ Mod.\ Phys.\ A {\bf 3}, 1615 (1988).
  doi:10.1142/S0217751X88000710
  
\bibitem{Amati:1987wq} 
  D.~Amati, M.~Ciafaloni and G.~Veneziano,
  ``Superstring Collisions at Planckian Energies,''
  Phys.\ Lett.\ B {\bf 197}, 81 (1987).
  doi:10.1016/0370-2693(87)90346-7
  
\bibitem{Veneziano:2004er} 
  G.~Veneziano,
  ``String-theoretic unitary S-matrix at the threshold of black-hole production,''
  JHEP {\bf 0411}, 001 (2004)
  doi:10.1088/1126-6708/2004/11/001
  [hep-th/0410166].
  
\bibitem{DAppollonio:2013okd} 
  G.~D'Appollonio, P.~Di Vecchia, R.~Russo and G.~Veneziano,
  ``The leading eikonal operator in string-brane scattering at high energy,''
  Springer Proc.\ Phys.\  {\bf 153}, 145 (2014)
  doi:10.1007/978-3-319-03774-5\_8
  [arXiv:1310.4478 [hep-th]].

\bibitem{Giddings:2006vu} 
  S.~B.~Giddings,
  ``Locality in quantum gravity and string theory,''
  Phys.\ Rev.\ D {\bf 74}, 106006 (2006)
  doi:10.1103/PhysRevD.74.106006
  [hep-th/0604072].
  
\bibitem{Giddings:2007bw} 
  S.~B.~Giddings, D.~J.~Gross and A.~Maharana,
  ``Gravitational effects in ultrahigh-energy string scattering,''
  Phys.\ Rev.\ D {\bf 77}, 046001 (2008)
  doi:10.1103/PhysRevD.77.046001
  [arXiv:0705.1816 [hep-th]].
  
\bibitem{Almheiri:2012rt} 
  A.~Almheiri, D.~Marolf, J.~Polchinski and J.~Sully,
  ``Black Holes: Complementarity or Firewalls?,''
  JHEP {\bf 1302}, 062 (2013)
  doi:10.1007/JHEP02(2013)062
  [arXiv:1207.3123 [hep-th]].
  
\bibitem{Almheiri:2013hfa} 
  A.~Almheiri, D.~Marolf, J.~Polchinski, D.~Stanford and J.~Sully,
  ``An Apologia for Firewalls,''
  JHEP {\bf 1309}, 018 (2013)
  doi:10.1007/JHEP09(2013)018
  [arXiv:1304.6483 [hep-th]].
  
\bibitem{Emparan:1999gf} 
  R.~Emparan,
  ``AdS / CFT duals of topological black holes and the entropy of zero energy states,''
  JHEP {\bf 9906}, 036 (1999)
  doi:10.1088/1126-6708/1999/06/036
  [hep-th/9906040].
    
\bibitem{Banados:1992wn} 
  M.~Banados, C.~Teitelboim and J.~Zanelli,
  ``The Black hole in three-dimensional space-time,''
  Phys.\ Rev.\ Lett.\  {\bf 69}, 1849 (1992)
  doi:10.1103/PhysRevLett.69.1849
  [hep-th/9204099].
      
\bibitem{Banados:1992gq} 
  M.~Banados, M.~Henneaux, C.~Teitelboim and J.~Zanelli,
  ``Geometry of the (2+1) black hole,''
  Phys.\ Rev.\ D {\bf 48}, 1506 (1993)
  Erratum: [Phys.\ Rev.\ D {\bf 88}, 069902 (2013)]
  doi:10.1103/PhysRevD.48.1506, 10.1103/PhysRevD.88.069902
  [gr-qc/9302012].
  
\bibitem{Horowitz:2009wm} 
  G.~Horowitz, A.~Lawrence and E.~Silverstein,
  ``Insightful D-branes,''
  JHEP {\bf 0907}, 057 (2009)
  doi:10.1088/1126-6708/2009/07/057
  [arXiv:0904.3922 [hep-th]].
  
\bibitem{Parikh:1999mf} 
  M.~K.~Parikh and F.~Wilczek,
  ``Hawking radiation as tunneling,''
  Phys.\ Rev.\ Lett.\  {\bf 85}, 5042 (2000)
  doi:10.1103/PhysRevLett.85.5042
  [hep-th/9907001].
  
\bibitem{Kraus:1994by} 
  P.~Kraus and F.~Wilczek,
  ``Selfinteraction correction to black hole radiance,''
  Nucl.\ Phys.\ B {\bf 433}, 403 (1995)
  doi:10.1016/0550-3213(94)00411-7
  [gr-qc/9408003].
  
\bibitem{Bachas:1995kx} 
  C.~Bachas,
  ``D-brane dynamics,''
  Phys.\ Lett.\ B {\bf 374}, 37 (1996)
  doi:10.1016/0370-2693(96)00238-9
  [hep-th/9511043].
 
\bibitem{Bachas:1992bh} 
  C.~Bachas and M.~Porrati,
  ``Pair creation of open strings in an electric field,''
  Phys.\ Lett.\ B {\bf 296}, 77 (1992)
  doi:10.1016/0370-2693(92)90806-F
  [hep-th/9209032].
  
\bibitem{Maldacena:2000hw} 
  J.~M.~Maldacena and H.~Ooguri,
  ``Strings in AdS(3) and SL(2,R) WZW model 1.: The Spectrum,''
  J.\ Math.\ Phys.\  {\bf 42}, 2929 (2001)
  doi:10.1063/1.1377273
  [hep-th/0001053].
   
\end{thebibliography}
\end{document}